\documentclass[a4paper]{jpconf}
\usepackage{graphicx}
\begin{document}
\title{Gauguin's questions in particle physics: \\
{\small Where are we coming from? What are we? Where are we going?}}

\author{John Ellis}

\address{Theory Division, Physics Department, CERN, CH-1211 Geneva 23, Switzerland}

\ead{John.Ellis@cern.ch}

\begin{abstract}
\vskip - 3in
\rightline{CERN-PH-TH/2007-210}
\rightline{October 2007}
\vskip 2.7in
Within particle physics itself, Gauguin's questions may be interpreted as:\\
P1 - What is the status of the Standard Model?\\
P2 - What physics may lie beyond the Standard Model?\\
P3 - What is the `Theory of Everything'?\\
Gauguin's questions may also asked within a cosmological context:\\
C1 - What were the early stages of the Big Bang?\\
C2 - What is the material content of the Universe today?\\
C3 - What is the future of the Universe?\\
In this talk I preview many of the topics to be discussed in the plenary sessions of this conference,
highlighting how they bear on these fundamental questions.
\end{abstract}

\section{Prologue}

\begin{figure}
\begin{center}
\includegraphics[width=2in,angle=270]{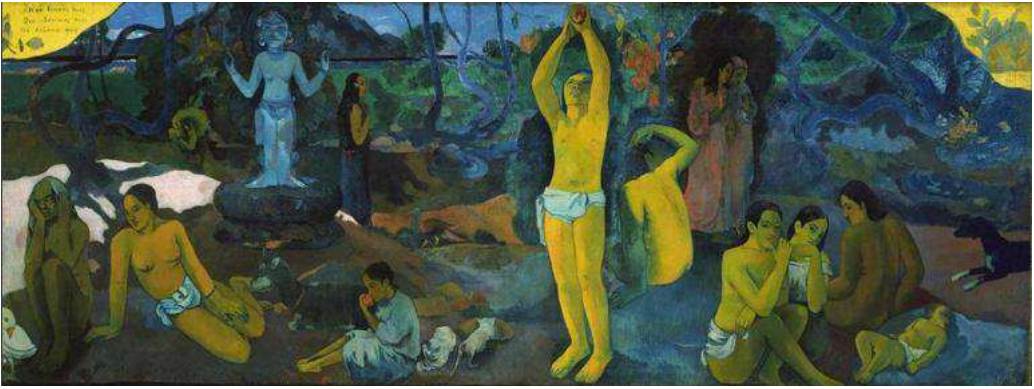}
\end{center}
\caption{\label{fig:Gauguin}
Gauguin's questions: D'o\`u venons-nous? Que sommes-nous? O\`u allons-nous? Where are we coming from? What are we? Where are we going?}
\end{figure}

As a research student, I stuck a copy of the Gauguin painting shown in Fig.~\ref{fig:Gauguin}
on the wall of my office, just to remind me why I was there. When the organizers asked me to
give an `inspirational' opening talk here, I opted to use Gauguin's questions to structure it. In the
following, I introduce many of the `Big Issues' in particle physics and cosmology, discussing
them in relation to the questions raised by Gauguin.

\section{What is the Status of the Standard Model? (P1)}

\subsection{Flavour and CP Violation}

\begin{figure}
\begin{center}
\includegraphics[width=3.2in]{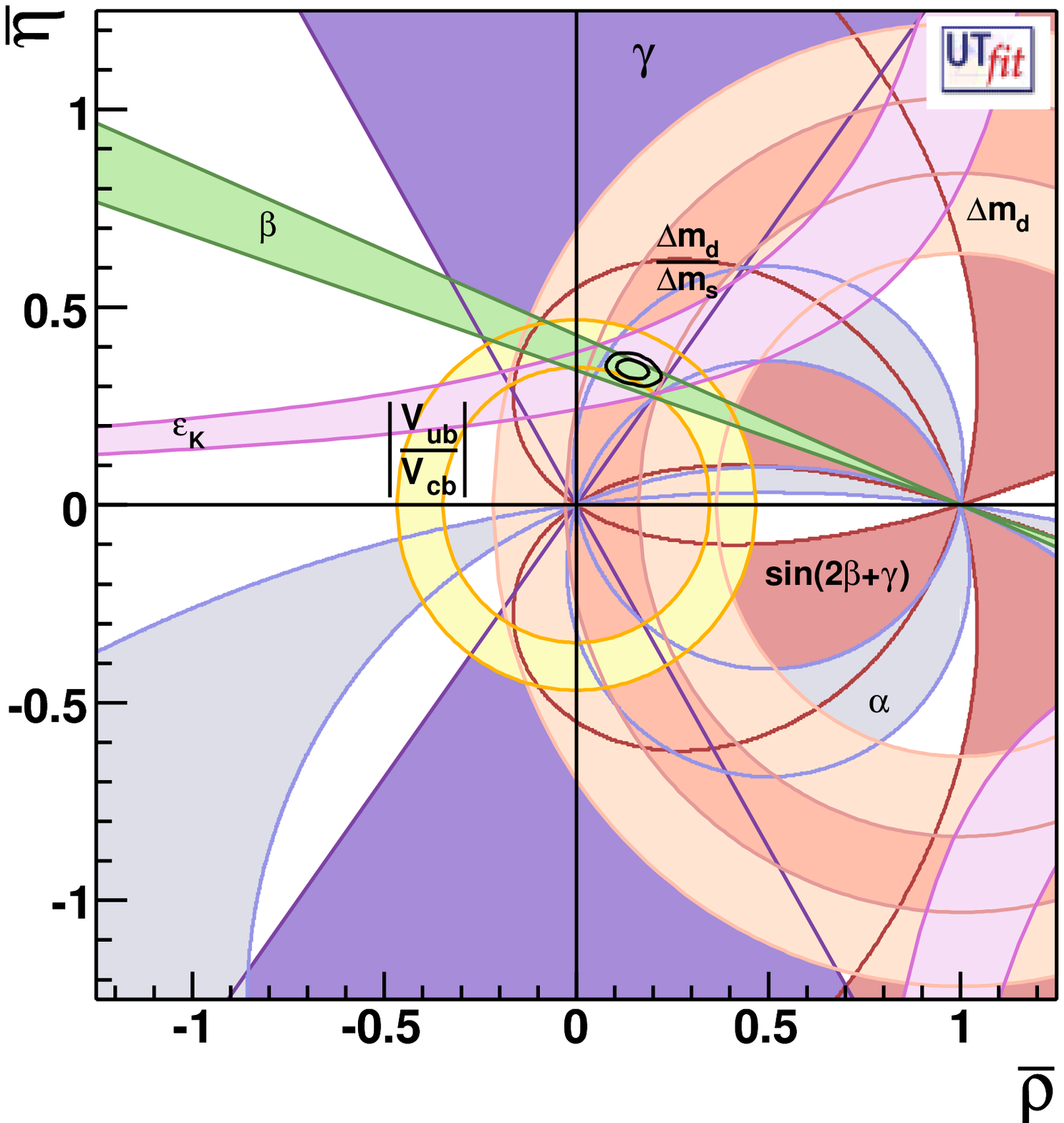}
\includegraphics[width=2.8in]{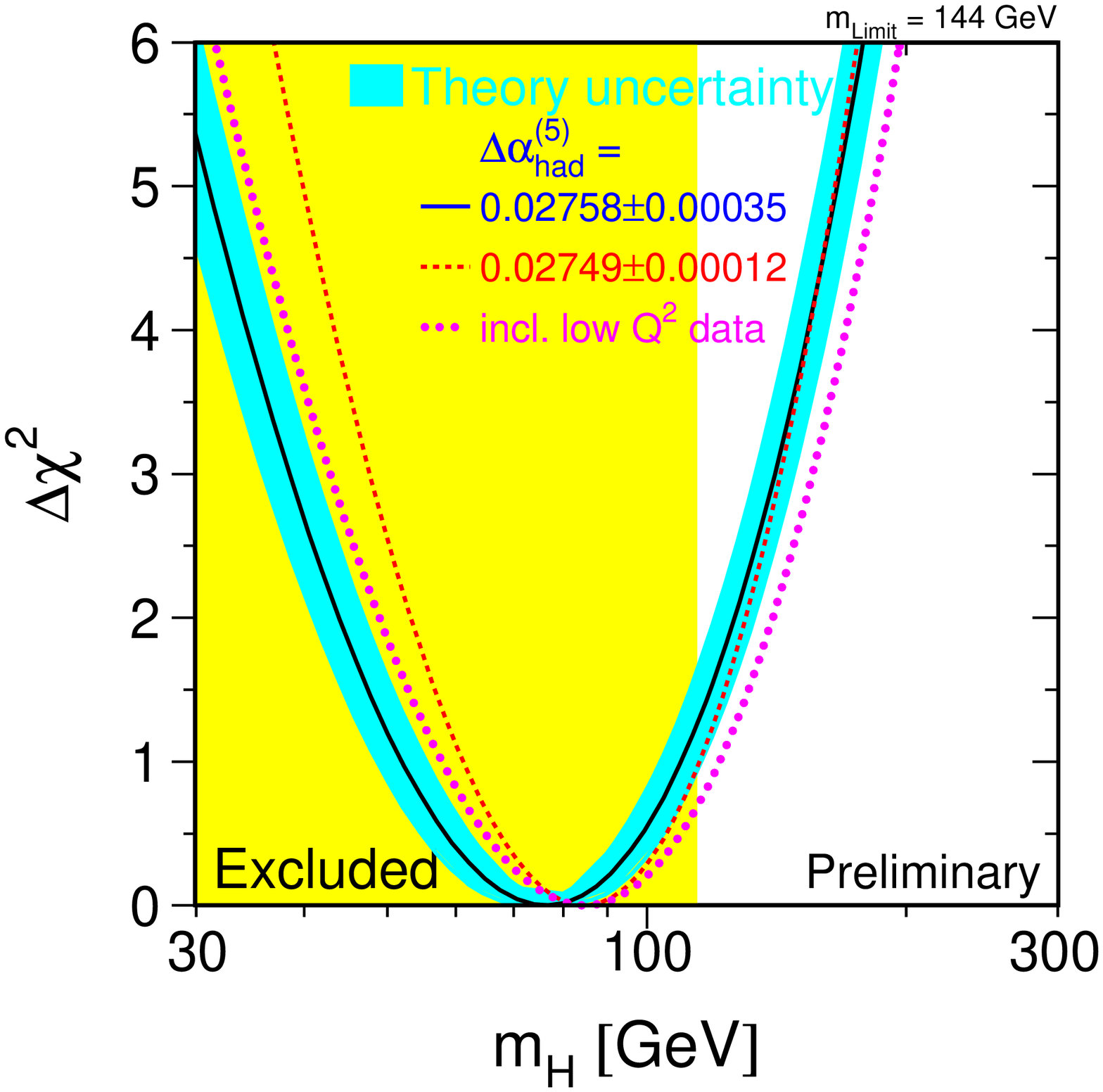}
\end{center}
\caption{\label{fig:CPX}Two pillars of the Standard Model: the left panel shows a global fit to the Kobayashi-Maskawa quark mixing parameters~\protect\cite{Flavour}, 
and the right panel shows $\chi^2$ as a
function of $m_H$ as obtained from a global fit to precision electroweak data~\protect\cite{EWWG}.}
\end{figure}

One of the most impressive developments in particle physics over the past few years has been
the continued success of the Kobayashi-Maskawa (KM) model of flavour and CP violation
within the Standard Model~\cite{KM}, worthily recognized by this year's EPS Prize for
High-Energy and Particle Physics. The torrent of data from the B factories and elsewhere,
including the Tevatron and fixed-target experiments~\cite{HFAG}, has been largely in
agreement with the KM model, as shown in the left panel of Fig.~\ref{fig:CPX}~\cite{Flavour}. 
There are no significant discrepancies
at present, though there some instances, e.g., $b \to s$ penguin processes, where
the data do not (yet) agree perfectly with the KM model. It is now clear that the KM
model explains most of the flavour and CP violation seen experimentally, and the
question is rather whether it needs to be supplemented by any additional flavour
physics beyond the KM model~\cite{Grossman}. KM is now the third pillar of the Standard Model,
joining QCD~\cite{QCD} and precision electroweak physics~\cite{Wyatt}.

\subsection{Precision Tests of the Electroweak Sector of the Standard Model}

For over a decade now, the precision electroweak data from LEP, the SLC, the
Tevatron and other experiments have not only been in very good overall
agreement with the Standard Model, but have also been providing
intriguing indications on the possible masses of unseen particles. The first
example was the top quark, whose measured mass $m_t = 170.9 \pm 1.9$~GeV~\cite{mt}
agrees to better than 10~\% with the value predicted within the Standard Model on
the basis of precision electroweak measurements. The second example is the
Higgs boson. Since the early 1990s~\cite{EFL}, the precision electroweak data have
been favouring, with increasing strength, a relatively light Higgs boson. The
one-$\sigma$ range currently indicated is~\cite{EWWG}
\begin{equation}
m_H \; = \; 76^{+33}_{-24} \; {\rm GeV},
\label{mhiggs}
\end{equation}
as shown in the right panel of Fig.~\ref{fig:CPX}.
The tendency towards a light Higgs boson has even been reinforced recently
by recent measurements of $m_t$ and $m_W$ at the Tevatron~\cite{Wyatt}. The
indication (\ref{mhiggs}) is still compatible with the direct lower limit
$m_H > 114$~GeV at the 15~\% level, and provides a tantalizing hint on the possible nature
of physics beyond the Standard Model. Broadly speaking, this would seem to favour
weakly-coupled models of new physics, such as supersymmetry.

\section{What Physics may lie Beyond the Standard Model?}

There is a standard list of fundamental open questions beyond the Standard Model.

\subsection{What is the Origin of Particle Masses?}

Are they due to a Higgs boson, as hypothesized within the Standard Model? 
If so, is the Higgs boson accompanied by some other physics? If not, what replaces the
Higgs boson? The good news is that, whatever the answers to these questions,
the puzzle is likely to be solved at some energy scale below 1~TeV~\cite{Gian}.\\
$\bullet$Why are there so many types of matter particles?\\
Related to this question is the mixing of the different flavours of quarks and leptons, and 
the mechanism for CP violation. This matter-antimatter difference is thought to be
responsible for the appearance of matter in the Universe today, and the absence of
antimatter. However, the KM mechanism within the Standard Model cannot, by itself,
explain the cosmological matter-antimatter asymmetry, adding urgency to the search
for flavour and CP violation beyond the Standard Model~\cite{Grossman}.\\
$\bullet$ Are the fundamental forces unified?\\
If so, in the simplest models this unification occurs only at some very high energy 
$\sim 10^{16}$~GeV. Physics at this scale cannot be probed directly at accelerators,
but possibly indirectly via measurements of particle masses and couplings, and looking
for unification relations between them. On the other hand, models of unification may be
probed more directly via neutrino physics~\cite{neutrino}.\\
$\bullet$ What is the quantum theory of gravity? (P3)\\
The two greatest successes of theoretical physics in the first half of the twentieth century
were quantum theory and general relativity. However, we still lack a full quantum
theory of gravity. The best candidate for such a theory may be (super)string theory~\cite{Lust},
which generously predicts extra space-time dimensions as well as supersymmetry, but at
what energy scale? Such a quantum theory of gravity would presumably be the 
long-sought `Theory of Everything' that
would answer the last of Gauguin's questions for fundamental physics.\\

The good news is that all of these fundamental open questions will be addressed by
the LHC: its energy should be ample for resolving the problem of mass, including the questions
whether there is a Higgs boson~\cite{LHCH} and/or supersymmetry~\cite{LHCsusy}, a dedicated
experiment will be examining matter-antimatter differences~\cite{LHCb}, models of unification could
be probed via measurements of sparticle masses and couplings, and string theory
might be probed via supersymmetry breaking, extra dimensions or even black hole
production and decay~\cite{Webberhole}. Thus, there is a lot of exciting new fundamental physics that
may be accessible to the LHC. However, some topics may only be accessible
indirectly to particle accelerators, so there is also an important role for astroparticle
experiments, which also bear on the cosmological aspects of Gauguin's questions.

\section{What were the Early Stages of the Big Bang? (C1)}

The Universe is remarkably isotropic on large distance scales, and apparently
also homogeneous. We have known for some 80 years that the Universe is expanding, 
and for a decade or so it has been apparent 
that this Big Bang expansion is even accelerating~\cite{SN}.
The cosmic microwave background (CMB) left over from the primordial electromagnetic
plasma is evidence that the visible part of the Universe was once thousands of time smaller 
and hotter than it is today, and the abundances of light elements indicate that the visible
Universe was once a million times hotter still. So much is certainly standard. Beyond
this Standard Big Bang Model of cosmology, it is believed
increasingly that the Universe expanded exponentially fast at some point in its much
earlier history, during an epoch of cosmological inflation~\cite{inflation} 
when quantum fluctuations appeared~\cite{flux}. 
These are thought to have led to the small anisotropies seen in the CMB
which, with the aid of cold dark matter, may have given rise to the structures seen in
the Universe today.  

\section{What is the Material Content of the Universe Today? (C2)}

\begin{figure}
\begin{center}
\includegraphics[width=5in]{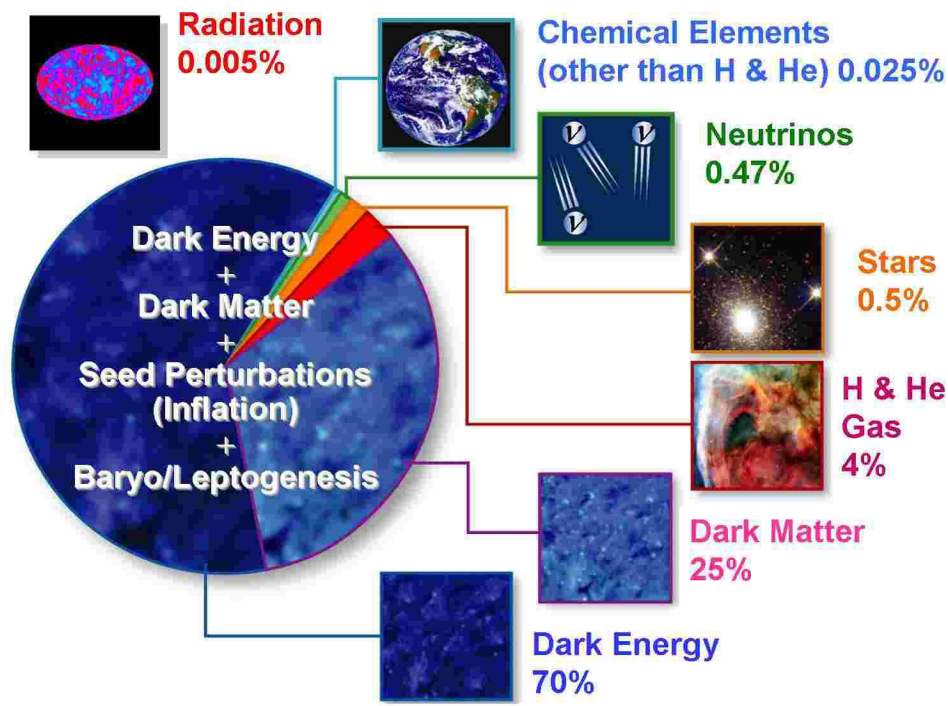}
\end{center}
\caption{\label{fig:pizza}
The toppings of the cosmic pizza~\protect\cite{Kolb}.}
\end{figure}

The material topping of the resulting cosmic pizza has a 
very strange recipe~\cite{Astier}. As shown in Fig.~\ref{fig:pizza}, the visible matter
makes up about 4~\% of its total energy density, presumably more than neutrinos,
which, in turn, contribute more than the CMB. Much more important are unidentified
contributions to the energy density, namely the cold dark matter that contributes
$\sim 25$~\%, with the rest of the energy density (some 70~\%) being provided by the 
so-called dark energy, which is not associated with matter at all, but in present
even in `empty' space.

This remarkable `concordance model' is prompted by a number of convergent
astrophysical and cosmological observations~\cite{triangle}. However, it begs a number of
fundamental open cosmological questions:\\
$\bullet$ Why is the Universe so big and old?\\
The Universe is almost 14 billion years old, and many billions of light-years across, at least. 
Why do we live in such an atypical cosmological solution of Einstein's equations, whose only
dimensional parameter corresponds to a time scale $\sim 10^{-43}$~s and a length
scale $\sim 10^{-33}$~s?\\
$\bullet$ Why is the geometry of the Universe so very nearly Euclidean?\\
The Universe is very nearly flat, with a density that is within a \% or so of the
critical density for a truly flat Universe. Cosmological inflation~\cite{inflation} would answer
this and the previous question, but how can we test this theory?\\
$\bullet$ Where did the matter in the Universe come from?\\
The Universe contains about a billion photons for every proton. Why not more or
less? Why any at all? The origin of matter in the Universe might be explained by
a suitable CP-violating matter-antimatter asymmetry~\cite{Sakharov}, but this must lie beyond the
simple KM model.\\
$\bullet$ What provides the dark matter?\\
The betting is that takes the form of one or more varieties of weakly-interacting massive
particle (WIMP), such as the lightest supersymmetric particle (LSP)~\cite{EHNOS}
and/or the axion~\cite{axion}. Perhaps
so, but this cannot be proven by astrophysical and cosmological observations alone: the
nature of the dark matter could only be established by laboratory experiments, e.g., at the LHC
in the case of a WIMP such as the LSP.\\
$\bullet$ What is the nature of the dark energy? (C3)\\
If this is constant, as originally postulated by Einstein, then the Universe will not
only expand for ever, but will end in the `heat death' of an accelerating de Sitter phase.

This particular answer to Gauguin's last question for cosmologists may not be very inspiring,
and perhaps particle physics will provide us with a more exciting answer. What is
sure, though, is that we need particle physics if we are to obtain answers to any of
these fundamental cosmological questions. We now see how they may be
addressed by experiments at the LHC.

\section{The LHC (P2, C2)}

The LHC is designed primarily to collide protons with energies 7~TeV each with a luminosity
of $10^{34}$~cm$^{-2}$s$^{-1}$~\cite{Evans}, corresponding to a billion collisions per second, with the
ability alternatively to collide heavy nuclei with energies 5~TeV per nucleon. Its primary
physics targets will include the origin of mass, the nature of dark matter, the nature of the primordial
quark-gluon plasma thought to have filled the Universe when it was less than a microsecond
old, and matter-antimatter asymmetry. It is clear that many of these objectives are connected with
Gauguin's questions in cosmology, as well as his questions in particle physics.

However, looking for the interesting new physics in proton-proton collisions at the LHC
will be non-trivial. The interesting cross sections for new heavy particles will be of order
1/(TeV)$^2$, and many interesting new particles such as the Higgs boson have small
couplings ${\cal O}(\alpha^2)$, whereas the total cross section is of order 1/(100~MeV)$^2$.
Interesting events may occur at relative rates $\sim 10^{-12}$, comparable to looking for
a needle in some 100,000 haystacks.

\begin{figure}
\begin{center}
\includegraphics[width=3.5in]{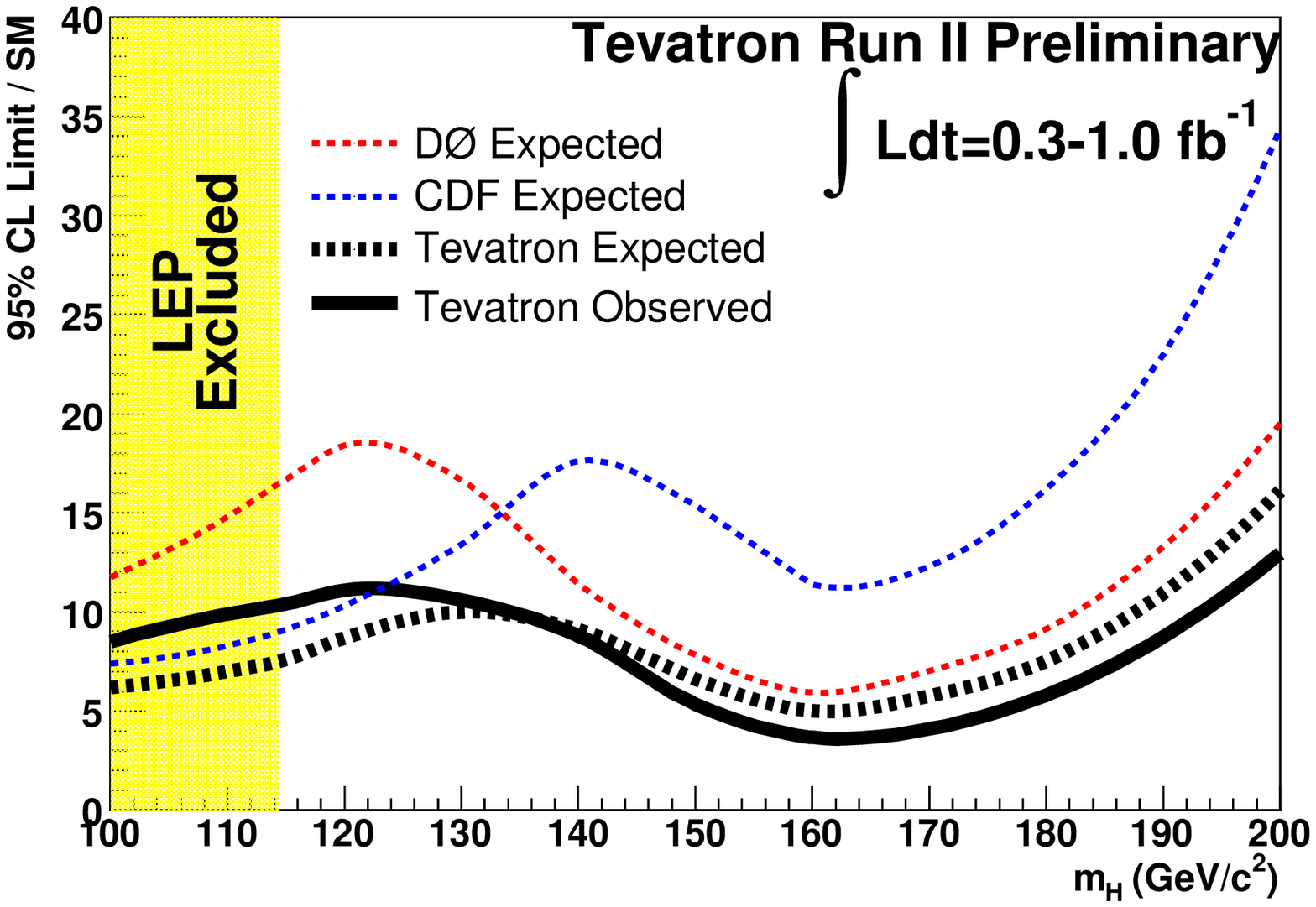}
\includegraphics[width=2.5in]{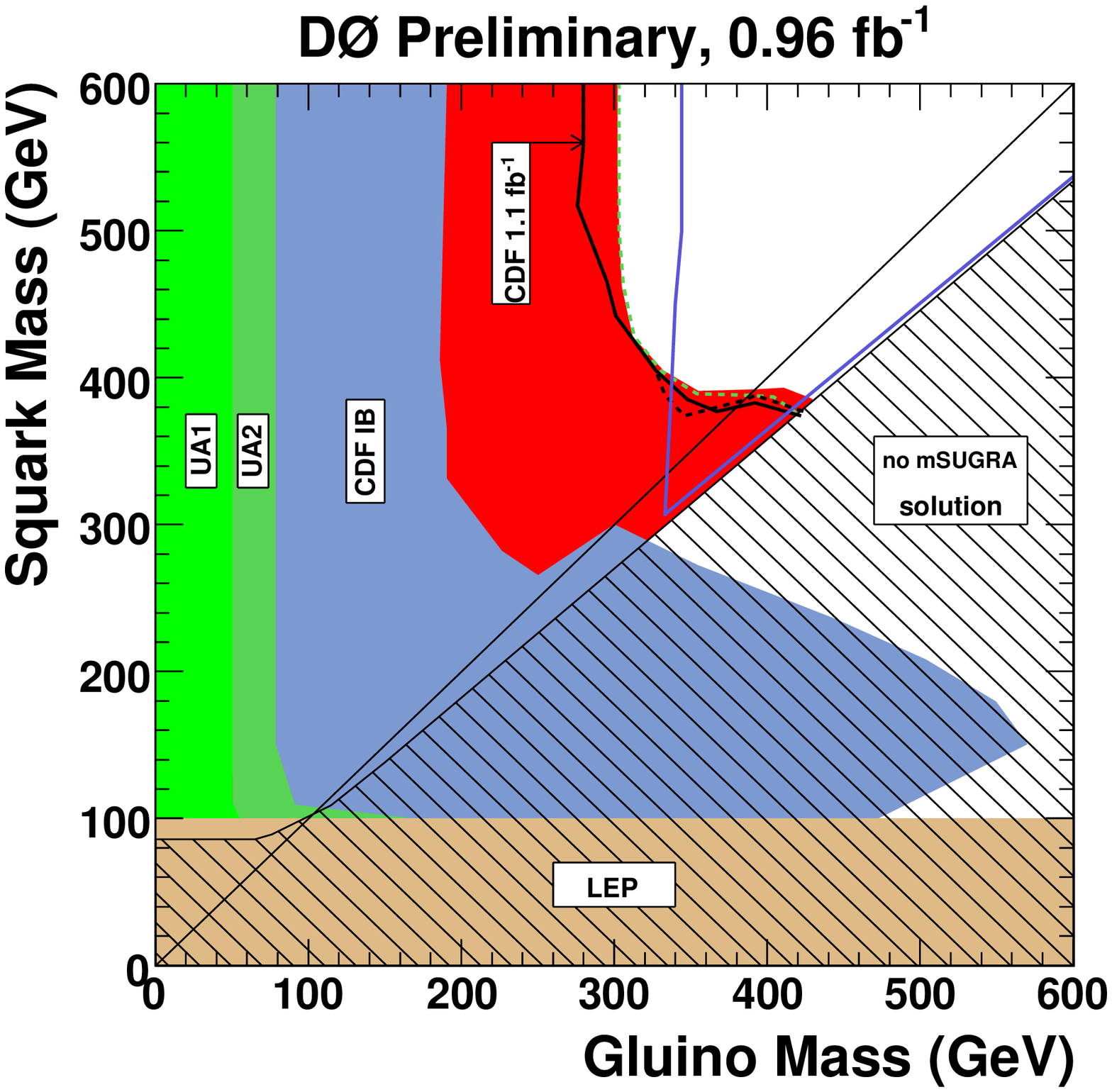}
\end{center}
\caption{\label{fig:Tevatron}The Tevatron is making good running in the search for the
Standard Model Higgs boson (left panel) and squarks and gluinos 
(right panel)~\protect\cite{Tevatron}.}
\end{figure}

The LHC's most anticipated discovery is surely that of the Higgs boson~\cite{Higgs}, but please do not
discount it! On the one hand, the Higgs boson may not exist~\cite{Higgsless} and, on the other, even a
Standard Model Higgs boson will be difficult to discover, let alone a non-standard one
with stage fright. A different threat is from the Higgs searches at the Tevatron, which are advancing ominously~\cite{Tevatron}. As seen in Fig.~\ref{fig:Tevatron},
they have already reached within an order of magnitude of the cross section
for a Standard Model Higgs boson with a mass just above the LEP lower limit, and are even
closer to the Standard Model cross section if $m_H \sim 160$~GeV. Moreover, not all
the available data have been fully evaluated, several more fb$^{-1}$ of data are
expected, and analysis techniques are constantly improving. A Standard Model Higgs
boson may well be within reach of the Tevatron.

\begin{figure}
\begin{center}
\includegraphics[width=6in]{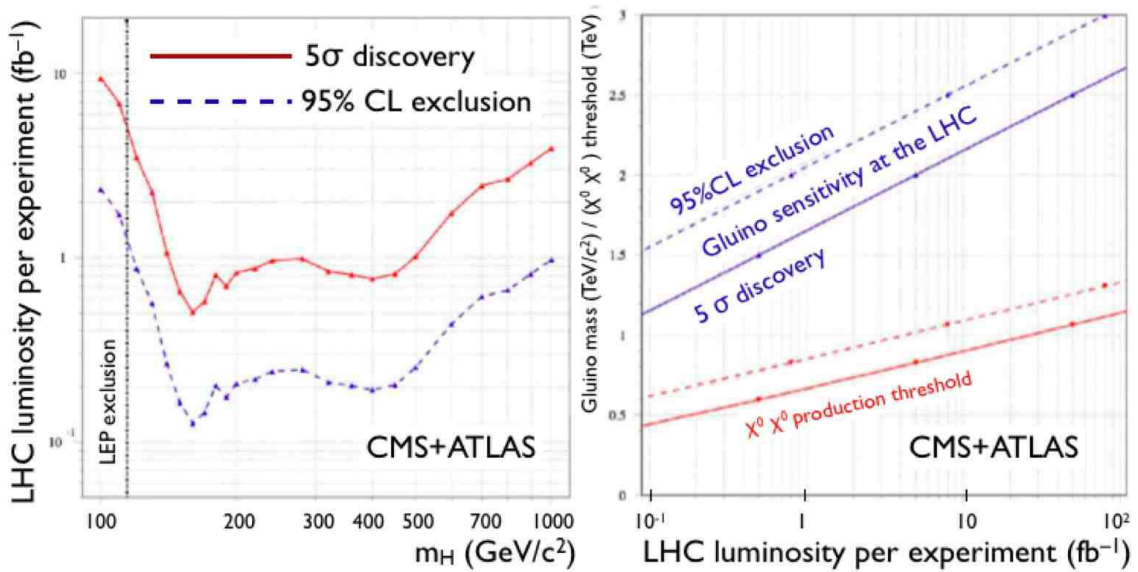}
\end{center}
\caption{\label{fig:LHCH}
The combined sensitivities of ATLAS and CMS to a
Standard Model Higgs boson (left), and the gluino (right), as a function of the
analyzed LHC luminosity. The right panel also shows the threshold for
sparticle pair production at a LC for the corresponding gluino mass, calculated
within the CMSSM~\protect\cite{POFPA}.}
\end{figure}

The Higgs signal at the LHC will be made up from several different signatures, such as
$H \to \gamma \gamma$, $H \to WW$, $H \to ZZ^* \to 4$ leptons and $H \to \tau \tau$, 
which must all be combined in order to extract the Higgs signal with high significance
in the early days of LHC running. This will require understanding very well details of the
responses of both the ATLAS and CMS detectors. As shown in Fig.~\ref{fig:LHCH},
estimates are that, by combining results from both
detectors, it should be possible to discover a Standard Model Higgs boson at the
5-$\sigma$ level with 5~fb$^{-1}$ of data each, while 1~fb$^{-1}$ of data each would
already enable a Standard Model Higgs boson to be excluded at the 95~\% level
over much of the possible mass range~\cite{POFPA}.

\section{Theorists getting Cold Feet (P2)}

\begin{figure}
\begin{center}
\includegraphics[width=5in]{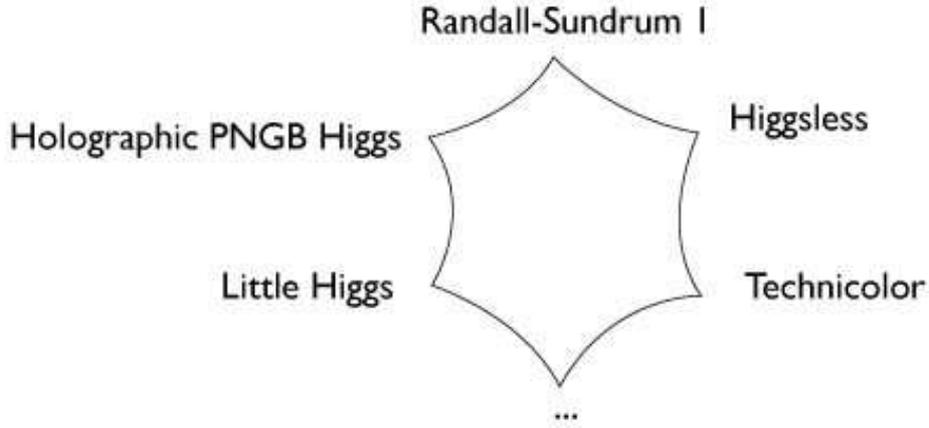}
\end{center}
\caption{\label{fig:Cheng}
An illustration of the space of possible alternatives to supersymmetry
at low energies~\protect\cite{Cheng}.}
\end{figure}

With the discovery of the Higgs boson at either the Tevatron or the LHC becoming
increasingly imminent, it seems that many theorists are getting cold feet, or
at least hedging their bets by dusting off alternative theories~\cite{Cheng}, many of
which are related, as shown in Fig.~\ref{fig:Cheng}. \\
$\bullet$
Perhaps the
elementary Higgs boson of the Standard Model should be replaced by some
composite alternative? \\
(Un)fortunately, at least simple technicolour models of this type run into
conflict with the precision electroweak data discussed earlier. \\
$\bullet$
So,  perhaps the
interpretation of the electroweak data is at fault~\cite{Chanowitz}? \\
There are issues with the mutual
consistency of the measurements, so perhaps some should be discarded, or perhaps
some new physics should be invoked to reconcile them? In either case, the usual
prediction of the mass of the Standard Model Higgs boson would be invalidated.\\
$\bullet$
Alternatively, perhaps the Standard Model Higgs boson is the right hypothesis,
but perhaps it is supplemented by higher-dimensional operators that allow the
mass range expected for the Higgs boson to be extended beyond (\ref{mhiggs}),
by opening up a corridor to heavier higher Higgs masses~\cite{Barbieri}? \\
$\bullet$
Another interesting
new class of Higgs scenarios comprises the so-called Little Higgs models~\cite{LHiggs}, 
see Fig.~\ref{fig:Cheng}, which
offer novelties such as an extra `top-likeÕ quark, gauge bosons,  and exotic Higgs
bosons. \\
$\bullet$
Finally, there are the `nightmare' scenarios offered by Higgsless models~\cite{Higgsless}, 
also seen in Fig.~\ref{fig:Cheng}.\\
In their original four-dimensional versions, these predicted strong $WW$ scattering
at the TeV scale which led, via loop effects, to incompatibilities with the precision
electroweak data. In order to mitigate this risk, it was proposed to break the
electroweak symmetry by boundary conditions in an extra dimension, which
enabled strong $WW$ scattering to be delayed until $\sim 10$~TeV. Even so,
issues of compatibility with precision electroweak data persist. Moreover, and
rather encouragingly, even if
one could construct such a Higgsless scenario, the likelihood is that it would have
other observable signatures such as Kaluza-Klein modes associated with the extra
dimension.

There is some debate what would be the maximum conceivable disaster scenario
for the LHC, assuming that the accelerator and the detectors work as designed. Some
might say that a Higgsless scenario would be disastrous, and it might indeed be
difficult to explain to funding agencies and politicians. On the other hand, it could be
a very promising avenue into extra space dimensions, which would arguably be more
exciting then `only' finding a Standard Model Higgs boson. A much duller scenario
might be to discover a Standard Model Higgs boson with a mass $\sim 180$~GeV,
in which case the renormalization group would permit you to run the Standard Model
parameters all the way to the Planck mass, and there might be no new physics before
quantum gravity sets in~\cite{boring}.

\section{The Stakes in the Higgs Search (P2)}

\begin{figure}
\begin{center}
\includegraphics[width=3in]{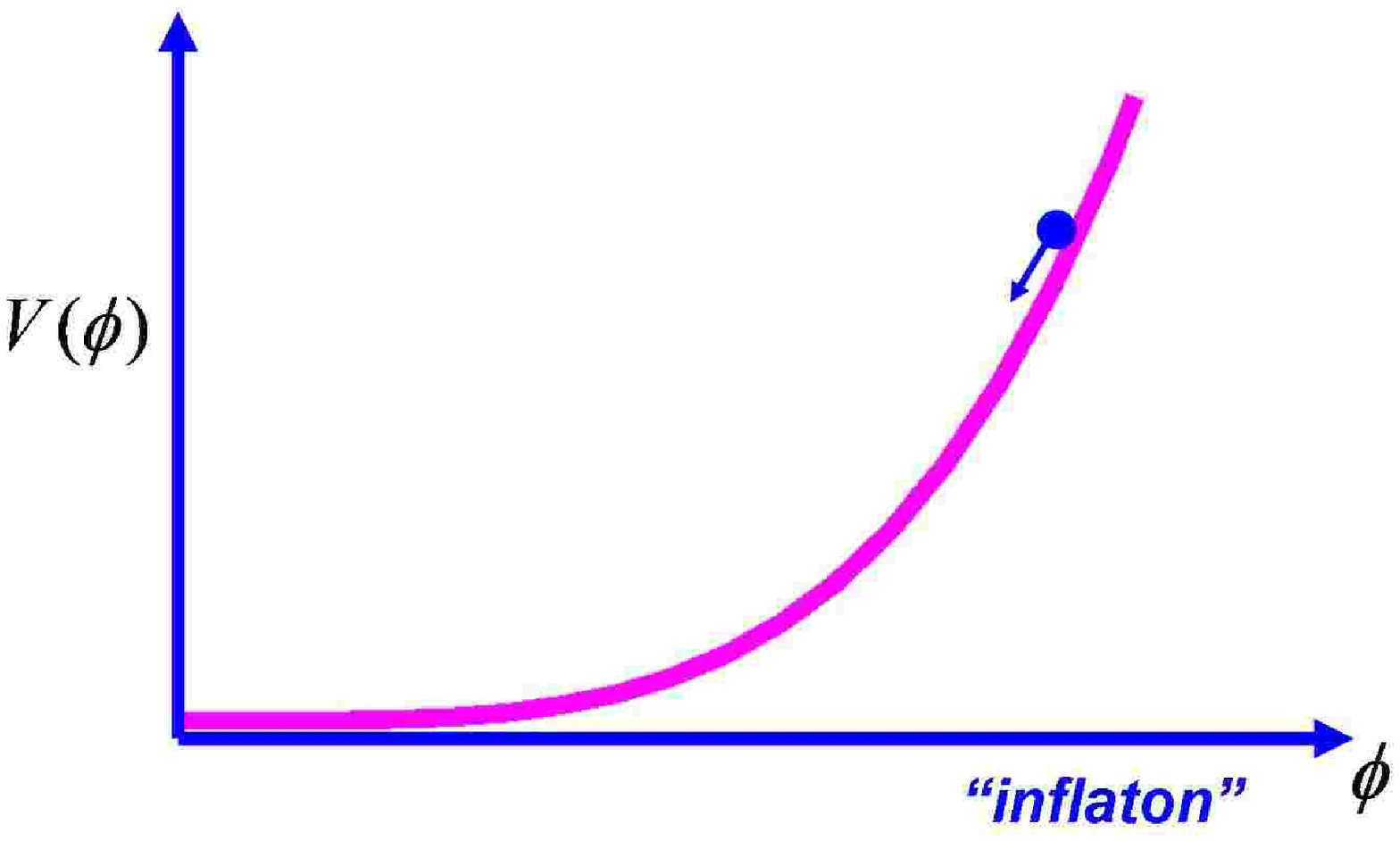}
\includegraphics[width=3in]{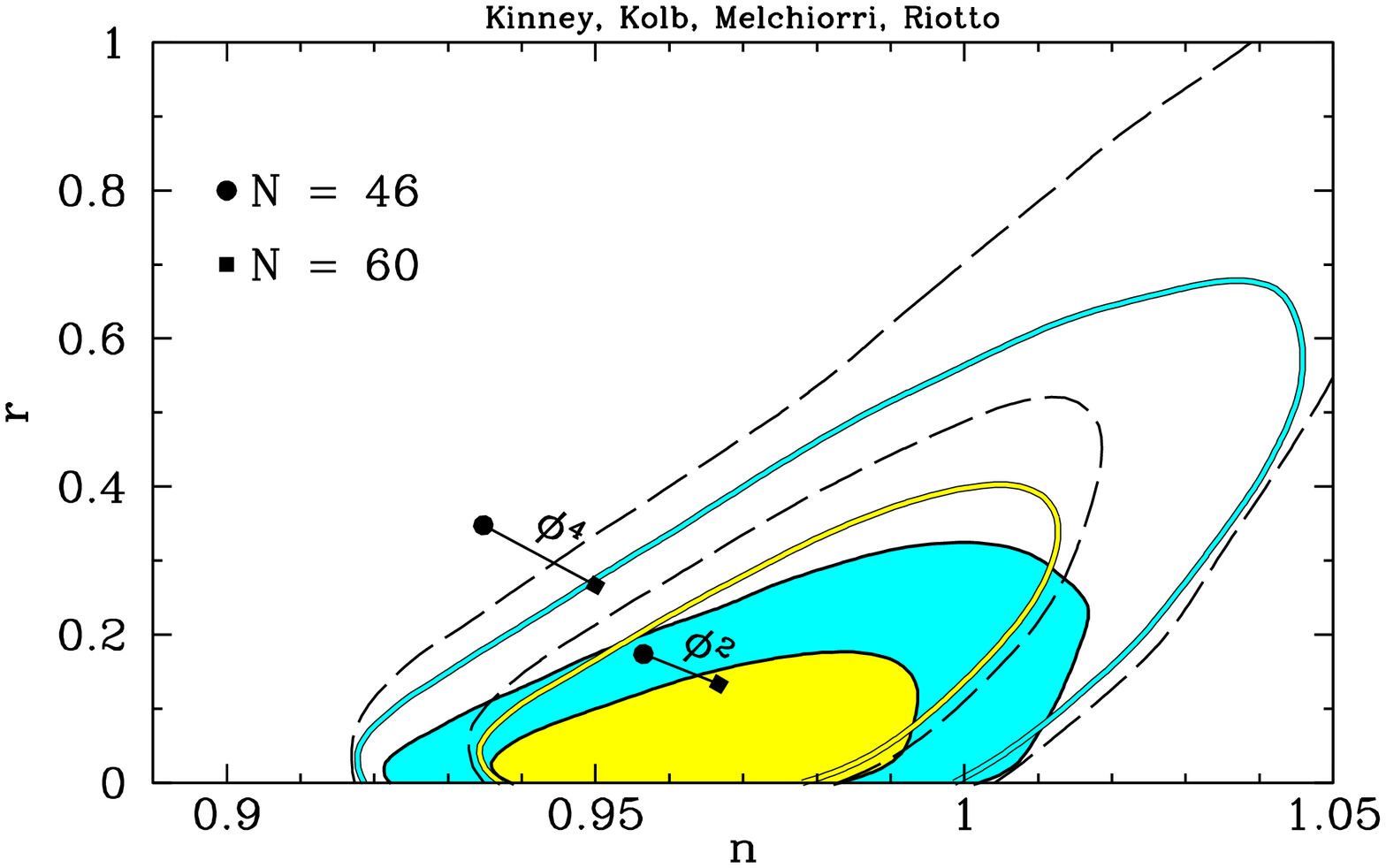}
\end{center}
\caption{\label{fig:inflaton}
Left panel: Inflation is driven classically by the vacuum energy associated with
the potential $V(\phi)$ of a scalar inflaton field $\phi$~\protect\cite{inflation}.
Right panel: Quantum fluctuations in $\phi$ would produce density perturbations~\protect\cite{flux},
and present measurements by WMAP {\it et al.} favour marginally a potential
$V \sim \phi^2$~\protect\cite{KKMR}.}
\end{figure}

There should be no misunderstanding: the stakes in the Higgs search are very high.
Within fundamental physics, issues are the manner in which gauge symmetry is
broken, and whether there is any elementary scalar field. However, the stakes for
cosmology are also very high. An elementary Higgs boson would have caused a
phase transition in the early Universe, when it was $\sim 10^{-12}$~s old (C1). It might,
at that epoch, have generated the matter in the Universe via electroweak
barygenesis (C2). Further back in the history of the Universe, a related inflaton~\cite{inflation} might
have expanded the Universe exponentially when it was $\sim 10^{-35}$~s old (C1, C2),
as illustrated in Fig.~\ref{fig:inflaton}.
Coming back to the present, naively the Higgs boson of the Standard Model
would contribute a factor $\sim 10^{56}$ too much to the present-day dark energy (C3),
apparently requiring some `miraculously' fine-tuned cancellation. Cosmologists
should be as interested as particle physicists in the d\'enouement of the Higgs saga.

\section{The LHC of Cosmology? (C1, C2)}

In parallel with the operation of the LHC, cosmologists will have the Planck satellite,
a new tool for analyzing the CMB that has similar potential for exploration within and
beyond the Standard Model of cosmology~\cite{Planck}. Its probes of the spectrum of primordial
density perturbations up to very high multipoles will provide stringent 
consistency tests of the inflationary idea illustrated in Fig.~\ref{fig:inflaton}. 
Planck measurements should also be
able to distinguish between specific models and yield detailed information about the 
effective inflationary
potential: is $V \sim \phi^n, e^{- \lambda \phi}$ or ...? Planck measurements will
enable the inflationary potential to be reconstructed over some range of values of
the inflaton field $\phi$. Measurements of the spectral index $n_s$ of the scalar
component of the perturbations, and the ratio $r$ of tensor perturbations to
scalar perturbations may discriminate between, e.g., a
quadratic potential $V \sim \phi^2$ and a quartic potential $V \sim \phi^4$. There is
already some preference for the simplest $\phi^2$~\cite{KKMR}, as seen in Fig.~\ref{fig:inflaton}.

Even if one accepts the inflationary paradigm, there are many options to
distinguish. Is the inflaton an elementary scalar field, or is it some
composite effective field that reflects more complex quantum-gravitational
(string?) dynamics? If there is an elementary inflaton field, what is it?
There is no candidate for the inflaton within the Standard
Model of particle physics: could it arise from some simple extension of the
Standard Model, or might it be some dilaton of string theory?

It is a challenge to relate the hypothetical inflaton to some recognizable particle physics.
One of the simplest possibilities is that the inflaton is a spartner of one of
the singlet (right-handed) neutrinos in a seesaw model of neutrino masses~\cite{snuinf}.
This would have an effective $m^2 \phi^2$ potential, and the magnitude of the
density perturbations observed by COBE, WMAP et al. fix $m \sim 2 \times 10^{13}$~GeV,
which fits well within the expected range of singlet neutrino masses. The decay of the
singlet neutrino could then generate the baryon asymmetry without further ado~\cite{ERY}.

\section{The Higgs Boson and Vacuum Energy (C3)}

As naively written: $V(\phi) = - \mu^2 |\phi|^2 + \lambda |\phi|^4$, the
effective Higgs potential would have a negative value at its minimum, which
would be about 56 orders of magnitude larger than the physical value of the
dark energy density. If particle theorists think about it at all, they just add a
constant to the effective Higgs potential, so as to cancel this unwanted value
to the desired 56 decimal places. This is just one aspect of the fine-tuning
problem posed by the dark energy. There is also a contribution to the
vacuum energy from the QCD vacuum that is some 44 orders of magnitude
too large. fashionable models of grand unification provide some 110 orders
of magnitude too much vacuum energy, and a generic model of quantum
gravity is likely to provide some 120 orders of magnitude too much! It seems
clear that understanding the observed  magnitude of the dark energy is
going to require some new physics. At the moment, all the data are consistent
with constant vacuum energy at the present epoch, in which case the Universe will expand exponentially for the foreseeable future.

\section{Supersymmetry (P2)}

My personal favourite candidate for (part of) this new physics is supersymmetry
which, with some luck, could even be the LHC's first discovery. There are several
reasons for liking supersymmetry: it is intrinsically beautiful, it may help unify the
different fundamental interactions, it is (almost) an essential ingredient in string
theory, etc. However, there are four specific reasons why one might expect 
supersymmetry to appear around the TeV scale, and hence be accessible to the LHC.
One is the naturalness or hierarchy problem~\cite{hierarchy}, another is the unification of the gauge
couplings~\cite{GUTs}, another is the supersymmetric prediction of a light Higgs boson~\cite{ENOS} 
as preferred by the precision electroweak data, and another is that many supersymmetric
models predict the existence of cold dark matter with a density comparable to that
required by astrophysics and cosmology~\cite{EHNOS}.

Supersymmetry helps make the hierarchy of mass scales in physics more natural~\cite{hierarchy},
by cancelling the quadratic divergences that arise from individual loop diagrams
with fermions and bosons:
\begin{eqnarray}
\Delta m_H^2 & = & - \frac{y_f^2}{16 \pi^2} [2 \Lambda^2 + 6 m_f^2{\rm ln}(\Lambda/m_f) + ... ],
\nonumber \\
\Delta m_H^2 & = & - \frac{\lambda_S}{16 \pi^2} [\Lambda^2 - 2 m_S^2{\rm ln}(\Lambda/m_S) + ... ].
\end{eqnarray}
We see immediately that the leading quadratic divergences cancel if $\lambda_S = 2 y_f^2$,
which is exactly the relation imposed by supersymmetry. Even more remarkably, this
cancellation persists to all orders in perturbation theory, rendering a light Higgs boson
technically natural if the supersymmetric partners of the Standard Model particles weigh
less than $\sim 1$~TeV. Remarkably, if sparticles have masses in this range, within
range of the LHC, detailed calculations show that they would improve the
renormalization-group unification of the gauge couplings~\cite{GUTs}, and that the
lightest supersymmetric particle (LSP) could provide the cold dark matter advocated by
astrophysicists and cosmologists (C2)~\cite{EHNOS}.

\section{Direct Evidence for Cold Dark Matter (C2)}

\begin{figure}
\begin{center}
\includegraphics[width=3.3in]{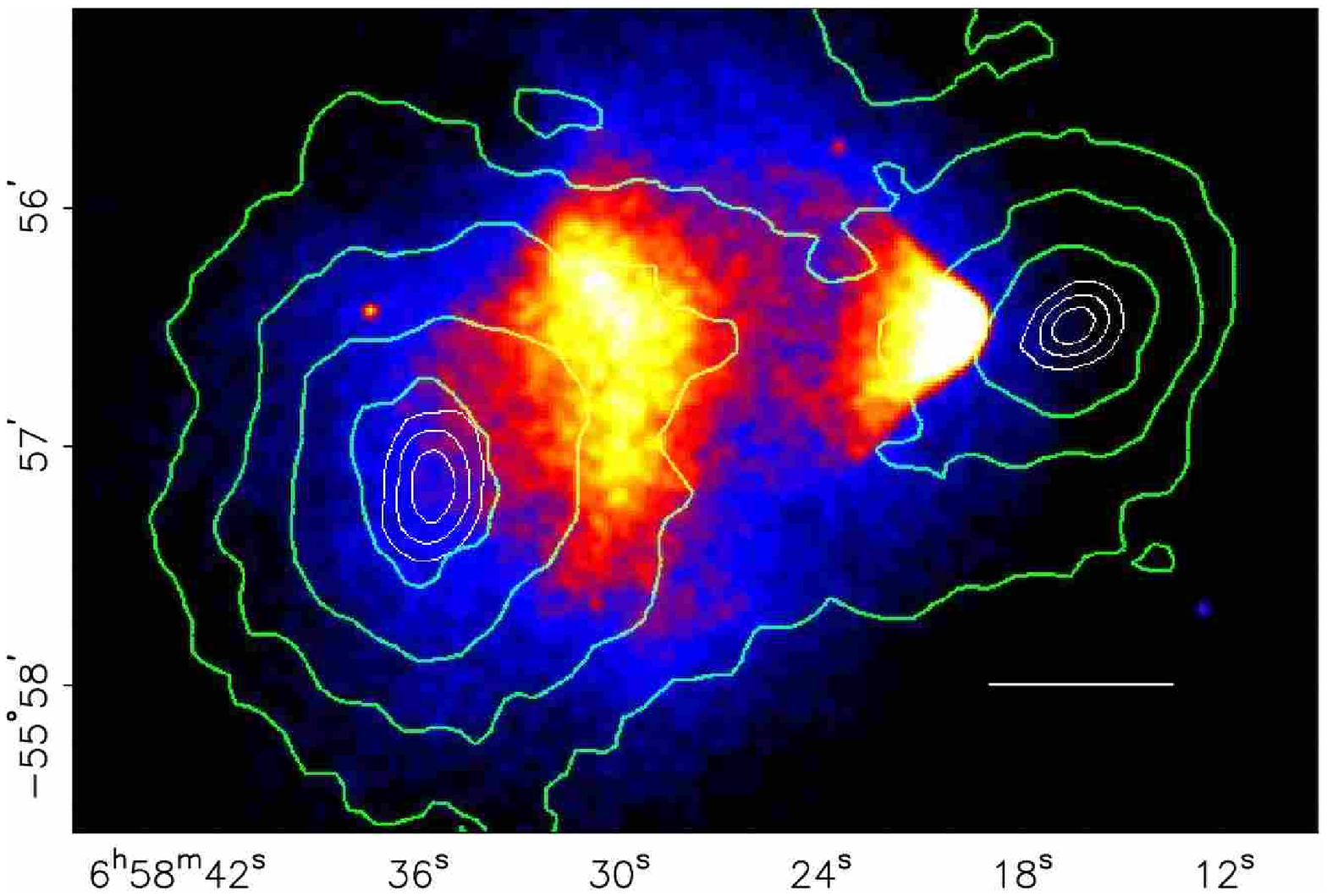}
\includegraphics[width=2.7in]{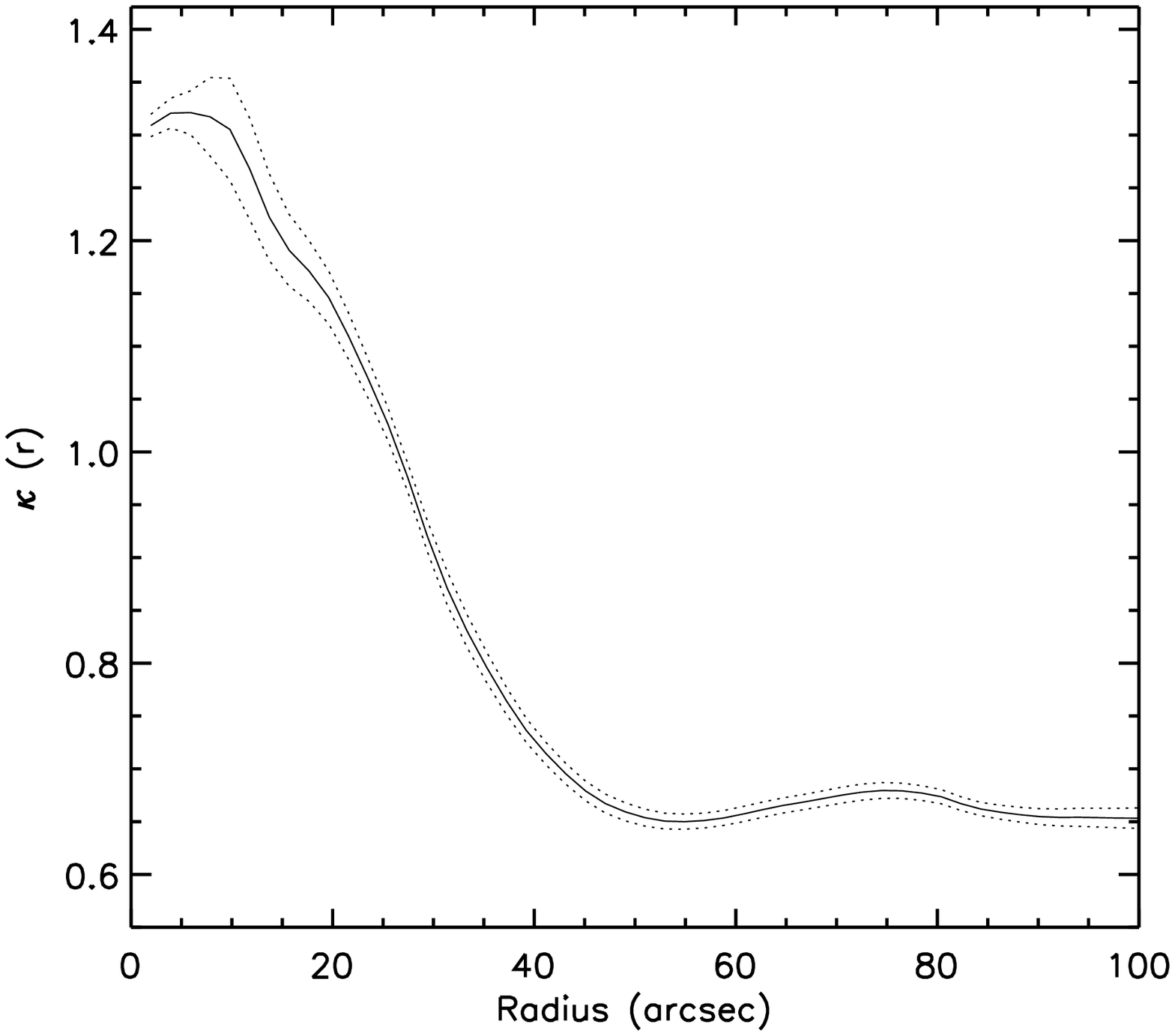}
\end{center}
\caption{\label{fig:collisions}Collisions between clusters of galaxies provide direct proof for weakly-interacting dark matter. In the left panel, two clusters have collided transverse to our line of sight,
and their dark matter cores have passed through each other essentially unscathed (contours),
whereas the associated gas seen between them has collided and heated up~\protect\cite{Clowe}. 
In the right panel,
two clusters have collided along our line of sight, and the radial distribution of the dark matter has
been perturbed, with a depletion at a radius $\sim 50$~arcsec and an enhancement at a
radius $\sim 75$~arcsec~\protect\cite{Jee}.}
\end{figure}

\begin{figure}
\begin{center}
\includegraphics[width=4in,angle=270]{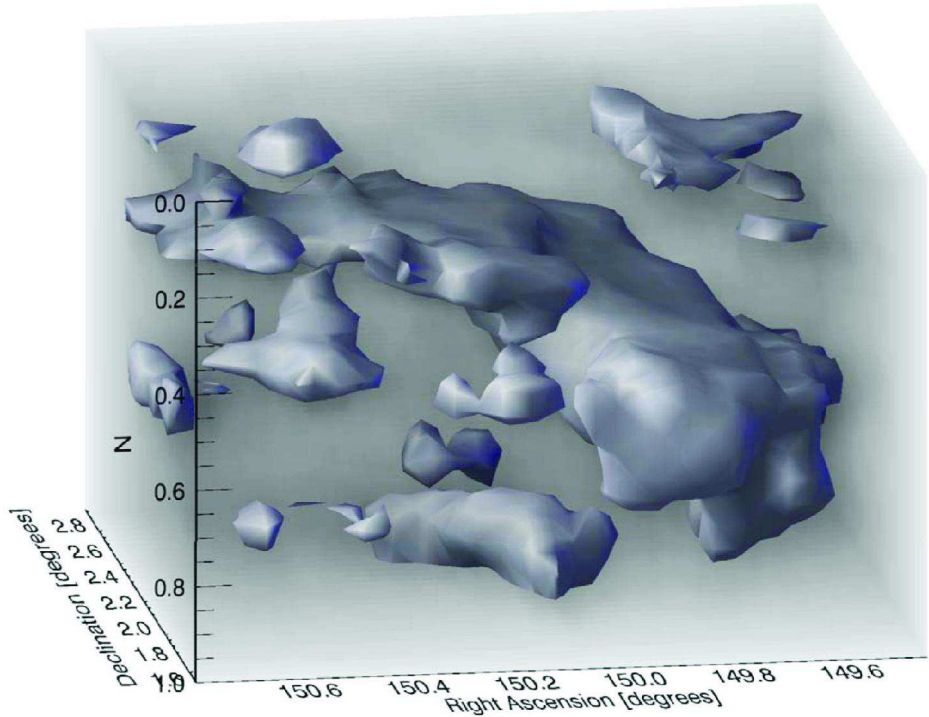}
\end{center}
\caption{\label{fig:scaffolding}A three-dimensional reconstruction of the scaffolding of
dark matter, as determined using weak lensing of light at different wavelengths,
originating on average from different redshifts~\protect\cite{scaffolding}.}
\end{figure}

In the past year, several dramatic pieces of direct evidence for collisionless dark matter
have emerged. For example, the results of a collision between two clusters of galaxies
moving across our line of sight
have been observed, as shown in the left panel of Fig.~\ref{fig:collisions}~\cite{Clowe}. 
Weak lensing of background objects shows that the dark matter
haloes in which the clusters are embedded have passed through each other essentially
without perturbation, whereas the gas clouds in the cluster have collided and heated up,
presumably because the gas molecules have much stronger interactions than the
dark matter particles. Another example is of the collision of two clusters along our
line of sight, shown in the right panel of Fig.~\ref{fig:collisions}~\cite{Jee}. 
In this case, a small perturbation is seen in the radial distribution of the cold 
dark matter, consistent with the expected tidal disruption expected from
gravitational interactions between the dark matter particles (cf, the `beam disruption
parameter' familiar at linear colliders). Thirdly, obervations of
weak lensing at different wavelengths (and hence coming on average from different
redshifts) has made possible a three-dimensional
deconstruction of the cold dark matter distribution, revealing a large scaffolding to which
visible matter is attached, shown in Fig.~\ref{fig:scaffolding}~\cite{scaffolding}.

\section{Constraints on Supersymmetry (P2)}

There are direct limits on sparticle masses from their absences at  LEP and the Tevatron,
and indirect constraints from the LEP lower limit $m_h > 114$~GeV and from $B$
physics, including in particular measurements of $b \to s \gamma$ decay. One
possible indication of new physics at the TeV scale may be provided by the BNL
measurement of the anomalous magnetic moment of the muon~\cite{BNL}, that seems to
exhibit a three-$\sigma$ discrepancy with the Standard Model, though this is still
somewhat controversial~\cite{Wyatt}. The strongest constraint on (one combination of)
supersymmetric model parameters is provided by the density of cold dark matter:
$0.094 < \Omega_\chi h^2 < 0.124$~\cite{Kolb}, assuming that it is mainly composed of the
lightest neutralino $\chi$. This not the only possibility: presumably the LSP should
have neither strong nor electromagnetic interactions~\cite{EHNOS}, but there are other
candidates that also have these properties. The supersymmetric partners of the neutrinos have
been excluded by a combination of LEP and direct dark matter searches, but the LSP
might be the spartner of some particle beyond the Standard Model, such as the gravitino.

\begin{figure}
\begin{center}
\includegraphics[width=2.7in]{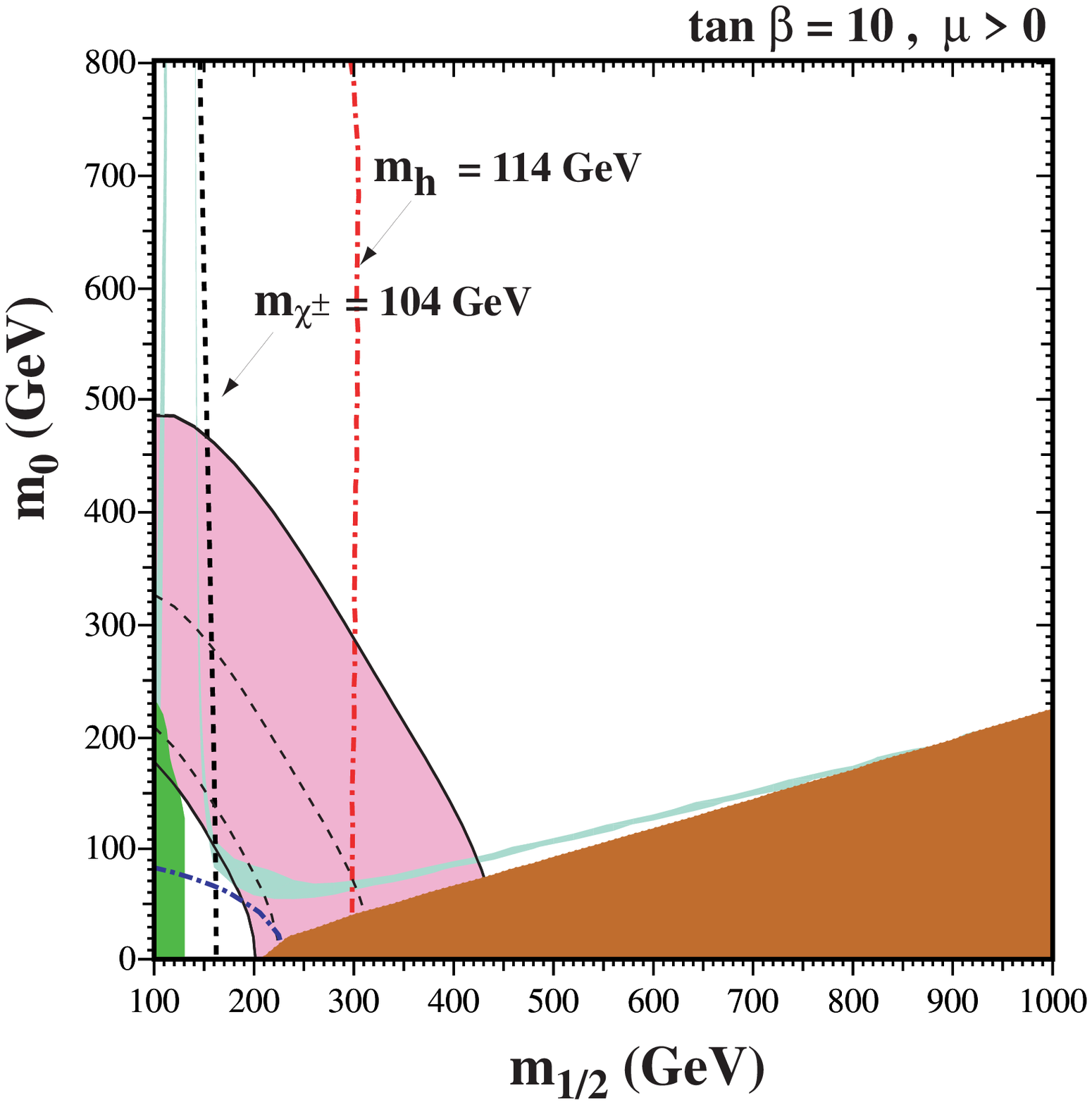}
\includegraphics[width=3.3in]{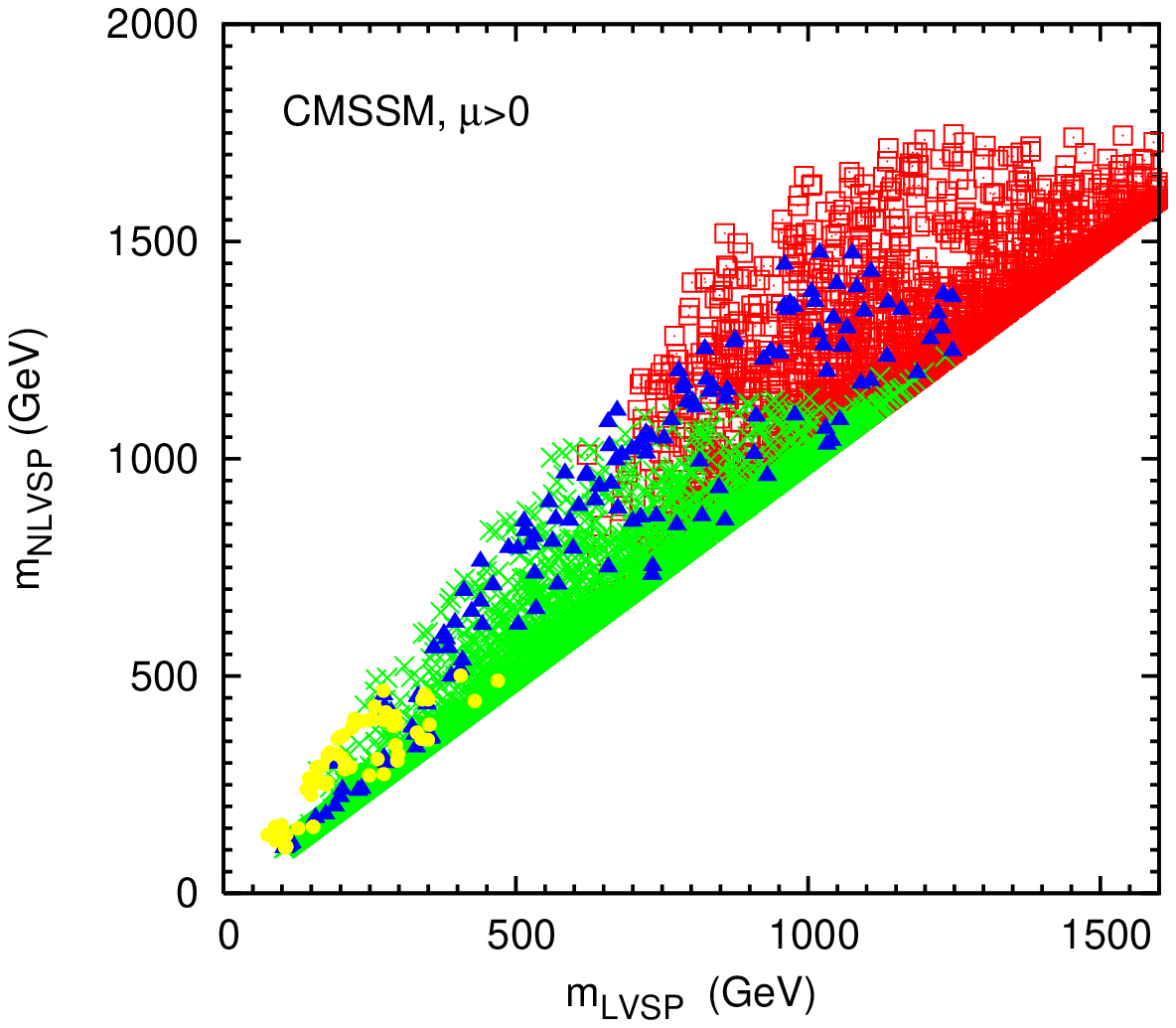}
\end{center}
\caption{\label{fig:CMSSM}Left panel: The $(m_{1/2}, m_0)$ plane in the CMSSM for
$\tan \beta = 10$, $\mu > 0$ and $A_0 = 0$~\protect\cite{EOSS}, 
incorporating the theoretical, experimental
and cosmological constraints described in the text. Right panel: The masses of the
lightest and next-to-lightest visible supersymmetric particles in a sampling
of CMSSM scenarios~\protect\cite{NLSP}. Also indicated are the scenarios
providing a suitable amount of cold dark matter (blue), those detectable at the LHC
(green) and those where the astrophysical dark matter might be detected
directly (yellow)~\protect\cite{NLSP}.}
\end{figure}

In a minimal supersymmetric model with universal soft supersymmetry-breaking
parameters, one may consider constraints in the $(m_{1/2}, m_0)$ plane, where
$m_{1/2}$ is the universal soft supersymmetry-breaking gaugino mass, 
and $m_0$ is the universal soft supersymmetry-breaking scalar mass. Regions
of this plane are excluded, in particular, by the requirement of a neutral LSP, by $m_h$, by
$b \to s \gamma$, by the dark matter density and (maybe) by $g_\mu - 2$. As shown
in the left panel of Fig.~\ref{fig:CMSSM}~\cite{EOSS}, the resulting
allowed regions are narrow strips near boundaries where $m_0 \sim 200$~GeV
and $m_\chi \sim m_{\tilde \tau_1}$, and where $m_0 > 1$~TeV and
electroweak symmetry breaking is no longer possible (which is disfavoured by $g_\mu - 2$).

Along these strips, $m_{1/2}$ may become quite large, and hence even the lightest
visible supersymmetric particle may become quite heavy, as shown in the
right panel of Fig.~\ref{fig:CMSSM}~\cite{NLSP}. Within a large sample (red points) of CMSSM
parameter choices, models yielding a suitable dark matter density are shown in blue.
Supersymmetric particles
would be detectable at the LHC in most (but not all) of these models (green points), whereas the
direct detection of supersymmetric dark matter might be possible only if the sparticles
are relatively light (yellow points).

\begin{figure}
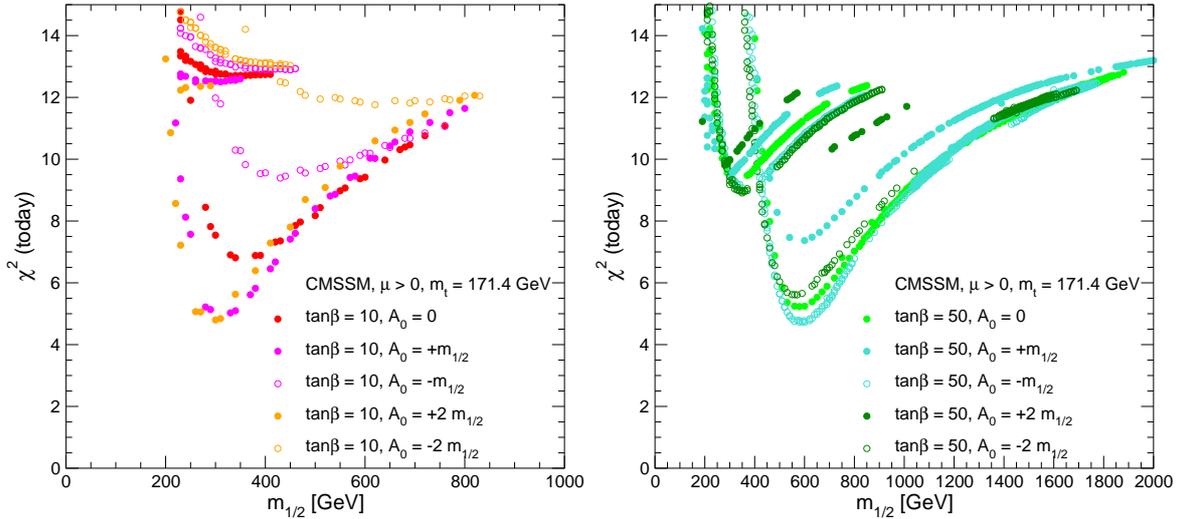

\begin{center}
\includegraphics[width=3in]{ehow5.CHI11a.1714.cl.eps}
\includegraphics[width=3in]{ehow5.CHI11b.1714.cl.eps}
\end{center}
\caption{\label{fig:EHOWW}
The combined $\chi^2$~function for electroweak precision
observables and $B$-physics observables, 
evaluated in the CMSSM for $\tan \beta = 10$ (left) and
$\tan \beta = 50$ (right) for various discrete values of $A_0$. We use
$m_t = 171.4 \pm 2.1$~GeV and $m_b(m_b) = 4.25 \pm 0.11$~GeV, and 
$m_0$ is chosen to yield the central value of the cold dark matter
density indicated by WMAP and other observations for the central values
of $m_t$ and $m_b(m_b)$~\protect\cite{EHOWW}.}
\end{figure}

As shown in Fig.~\ref{fig:EHOWW}~\cite{EHOWW},
a global fit to precision electroweak and $B$-decay observables indicates a
preference for relatively small values of $m_{1/2}$. This is due predominantly to
$g_\mu - 2$~\cite{BNL}, but there is some support from the measurements of $m_W$.
Correspondingly, the most likely value for the mass of the lightest supersymmetric 
Higgs boson is only slightly above the LEP lower limit.

\section{Searching for Supersymmetry (P2)}

The classic signature of sparticle production is the appearance of events with
missing energy-momentum carried away by invisible dark matter particles.
Studies indicate that such a signature should be observable at the LHC above
instrumental and Standard Model backgrounds. As shown in the right panel of
Fig.~\ref{fig:LHCH}~\cite{POFPA},
it is estimated that 0.1~fb$^{-1}$
of LHC luminosity would be sufficient to observe a gluino with a mass of 1.2~GeV
at the five-$\sigma$ level, or to exclude a gluino weighing $< 1.5$~TeV. The
discovery and exclusion reaches would extend to about 2.2 and 2.5~TeV, respectively
with 10~fb$^{-1}$ of LHC luminosity.

As also shown in the right panel of Fig.~\ref{fig:LHCH},
assuming universal input sparticle masses at the GUT scale, are the corresponding
thresholds for sparticle pair production at a linear $e^+ e^-$ collider would be
0.5 (0.6) or 0.8 (1.0) TeV~\cite{POFPA}. Hence, for example, if the LHC discovers the gluino
with 0.1~fb$^{-1}$, one may expect that sparticle pair production would be accessible
to a linear collider with centre-of-mass energy of 0.5~TeV, whereas if the LHC
does not discover the gluino even with 10~fb$^{-1}$, the $e^+ e^-$ sparticle
pair production threshold may be above 1~TeV. At least in such a simple model,
the LHC will tell us how much energy a linear collider would need to find supersymmetry.
If Nature is kind, and sparticles not only exist but also are quite light, it will be possible
to test directly unification of the gauge couplings and universality of the soft
supersymmetry-breaking scalar masses with high precision, in particular by
comparing measurements at the LHC and the ILC or CLIC.

\section{The Stakes in the SUSY Search (P2)}

The stakes in the search for supersymmetry are very high.
Supersymmetry would be a completely novel symmetry, never seen before
in Nature at the fundamental level. It would correspond, in a sense to a
novel `quantum' extension of the notion of space to `superspace'. It would
provide a circumstantial hint for string theory, which depends (almost)
on its existence. In addition to these fundamental insights, supersymmetry
would also stabilize the hierarchy of mass scales in physics, might
explain 90~\% of the matter in the Universe, and would facilitate
the unification of the fundamental forces. No wonder that so many theorists
are so enamoured of SUSY, and even some experimentalists! As seen in the right panel
of Fig.~\ref{fig:LHCH}, we might be lucky at the LHC and see supersymmetry quite quickly,
if the sparticles are light. Equally, however, we may need to be patient. There is not a 
complete guarantee that SUSY will
be found at the LHC, even if it does provide the dark matter, and 
a linear $e^+ e^-$ collider may need a centre-of-mass energy in the
multi-TeV range to be reasonably sure of observing even the lightest
visible supersymmetric particle~\cite{NLSP}.

\section{Search for a `Theory of Everything' (P3)}

The two greatest achievements of twentieth-century physics, namely
quantum mechanics and gravity, are still not combined. Moreover, we do
not have a unified description of gravity and the other fundamental
interactions, whose carrier particles have different spins. The `only
game in town' for solving these problems seems to be string theory~\cite{Lust}. As
already mentioned, strings seem (almost) to require supersymmetry,
and also revive the suggestion that there may be additional dimensions
of space.

In addition to its candidature as a `Theory of Everything', string theory
provides us with a tremendously powerful new theoretical toolbox that
goes beyond traditional quantum field theory in many ways. The insights
it provides have many other potential appliactions to more mundane problems,
such as perturbative calculations in QCD~\cite{pQCD} and the nature of the quark-gluon
plasma~\cite{QGP}.

The full enormity of the ambiguity in the string vacuum has sunk in only recently,
with numbers ${\cal O}(10^{500})$ being bandied about~\cite{landscape}. This ambiguity arises
because there are certainly millions and perhaps billions of consistent
compactifications of strings on manifolds in extra dimensions, and each of these
has dozens or hundreds of topological cycles through which there may be
topological fluxes taking any of dozens of values. Somewhere in this landscape of an
enormous number of string vacua, it is suggested there may be one with
a vacuum energy in the range indicated by the cosmology dark energy.
The question then arises how the Universe chooses which of these vacua.
One may also wonder whether, since nature apparently has the opportunity
to choose a small vacuum energy, perhaps it also chooses a small value
of $m_W$, and there is no need for supersymmetry to render the choice natural~\cite{nosusy}.

I was recently asked to give to young string theorists some introductory lectures
on LHC physics, and I peppered my lectures with the following ten questions
for string theorists:

$\bullet$ Can you derive (something like) the Standard Model from string, with
the appropriate SU(3)$\times$SU(2)$\times$U(1) gauge group, without too many
unseen additional particles, and does your derivation provide useful constraints
on the many arbitrary parameters of the Standard Model?

$\bullet$ Can you calculate the vacuum energy? Preferably, do not cop out 
by invoking the landscape!

$\bullet$ Are you able to live with a cosmological constant? Traditional string
theory is based on the existence of an $S$ matrix, but this does not exist in a de Sitter
space, as would be required if the vacuum energy is truly constant~\cite{Witten}. In such a case,
scattering is described by a superscattering $\$$ matrix~\cite{EMN}.

$\bullet$ Do you have any new ideas for avoiding the existence of a Higgs boson?
This would require developing	innovative strategies for gauge symmetry breaking, e.g.,
via boundary conditions in extra dimensions.
 
$\bullet$ Will your string-inspired techniques in perturbative QCD
enable better calculations of Higgs production and decay?
 
$\bullet$ Can you improve perturbative QCD techniques sufficiently to
improve calculations of the backgrounds to Higgs production and decay?
	
$\bullet$ Is supersymmetry really essential for string theory and, if so, are you
able to offer any hint on the scale of superymmetry breaking?

$\bullet$ What is the pattern of supersymmetry breaking and, in particular,
what are the relative values of $m_0$ and $m_{1/2}$, and are they universal?

$\bullet$ If they are universal, what is the scale where supersymmety-breaking 
parameters appear to be unified, and does supersymmetry breaking appear below the 
GUT/string scale?

$\bullet$ How large might extra dimensions be and, in particular,
is there any reason to expect them  below the GUT/string scale?

Answers to any of these questions would be a tremendous success for string theorists,
and help to quieten down the doubters.

\section{How Large could Extra Dimensions be? (P2)}

As already mentioned, one of the possibilities offered by string theory is that
there might be extra dimensions: how large could they be? When string theory
was originally proposed as a `Theory of Everything', it was imagined that all
the extra dimensions would be curled up on length scales comparable to the
Planck length $\sim 10^{-33}$~cm. However, then it was realized that string
unification could be achieved more easily if one of these dimensions was
somewhat smaller than the GUT scale~\cite{GUTsize}, and a number of scenarios with much
larger extra dimensions have been considered. For example, an extra
dimension of size $\sim 1$~TeV$^{-1}$ could help break supersymmetry~\cite{Ignatios}
and/or the electroweak gauge symmetry, an extra dimension of micron size
could help rewrite the hierarchy problem~\cite{ADD}, and even infinite extra dimensions are
allowed if they are warped appropriately~\cite{RS}.

In many of these scenarios, there are potential signals to be found at the LHC,
such as Kaluza-Klein excitations of gravitons, or missing energy `leaking' into
an extra dimension. The most spectacular possibility would occur if gravity becomes
strong at the TeV energy scale, in which case microscopic black holes might be 
produced at the LHC. These would be very unstable, decaying rapidly via
Hawking radiation into multiple jets, leptons and photons~\cite{Webberhole}.

\section{Conclusions}

These examples have shown that high-energy accelerators such as the LHC
are telescopes as well as microscopes. They are able to address Gauguin's
questions for cosmology and astrophysics, as well as in particle physics
itself. We do not know what the LHC will find, but surely Gauguin would be happy
with some of the answers it will provide.


\section*{References}
\begin{thebibliography}{99}

\bibitem{KM}
Kobayashi~M and Maskawa~T 1973
  {\it Prog.\ Theor.\ Phys.}  {\bf 49} 652
  
\bibitem{HFAG}
Barberio~E {\it et al}  [Heavy Flavor Averaging Group (HFAG)
                  Collaboration],
  arXiv:0704.3575 [hep-ex]
  
\bibitem{Flavour}
Bona~M {\it et al}  [UTfit Collaboration]
  arXiv:0707.0636 [hep-ph]
  
\bibitem{Grossman}
Grossman~Y, talk at this meeting

\bibitem{QCD}
Behnke~O, Nason~P, talks at this meeting

\bibitem{Wyatt}
T.~Wyatt, talk at this meeting

\bibitem{EWWG}
Grunewald~M~W
  arXiv:0710.2838 [hep-ex]
  
\bibitem{mt}
CDF Collaboration and D0 Collaborations
  arXiv:hep-ex/0703034
  
\bibitem{EFL}
Ellis~J~R, Fogli~G~L and Lisi~E 1993
  {\it Phys.\ Lett.}  B {\bf 318} 148
  
\bibitem{Gian}
Giudice~G~F, talk at this meeting
  arXiv:0710.3294 [hep-ph]
  
\bibitem{neutrino}
Brice~S, Lindner~M, talks at this meeting

\bibitem{Lust}
L\"ust~D, talk at this meeting

\bibitem{LHCH}
For a recent review, see:
Ruwiedel~C  [ATLAS Collaboration]
  arXiv:0710.1954 [hep-ph]
  
\bibitem{LHCsusy}
For a recent review, see:
Tytgat~M
  arXiv:0710.1013 [hep-ex]

\bibitem{LHCb}
LHCb Collaboration, {\tt http://lhcb.web.cern.ch/lhcb/}

\bibitem{Webberhole}
Harris~C~M, Richardson~P and Webber~B~R 2003
  {\it JHEP} {\bf 0308} 033
  [arXiv:hep-ph/0307305]
  
\bibitem{SN}
Riess~A~G {\it et al}  [Supernova Search Team Collaboration] 1998
  {\it Astron.\ J.} {\bf 116} 1009
  [arXiv:astro-ph/9805201];
  Perlmutter~S {\it et al}  [Supernova Cosmology Project Collaboration] 1999
  {\it Astrophys.\ J.}  {\bf 517} 565
  [arXiv:astro-ph/9812133]
  
\bibitem{inflation}
Guth~A~H 1981
  {\it Phys.\ Rev.}  D {\bf 23} 347;
Linde~A~D 1982
  {\it Phys.\ Lett.}  B {\bf 108} 389;
  Albrecht~A and Steinhardt~P~J 1982
  {\it Phys.\ Rev.\ Lett.} {\bf 48} 1220
  
\bibitem{flux}
Guth~A~H and Pi~S~Y 1982
  {\it Phys.\ Rev.\ Lett.}  {\bf 49} 1110;
  Bardeen~J~M, Steinhardt~P~J and Turner~M~S 1983
  {\it Phys.\ Rev.}  D {\bf 28} 679;
  Hawking~S~W and Moss~I~G 1983
  {\it Nucl.\ Phys.}  B {\bf 224} 180

\bibitem{Astier}
Astier~P, talk at this meeting

\bibitem{Kolb}
For a recent review, see: Kolb~E~W
  arXiv:0709.3102 [astro-ph]
  
\bibitem{triangle}
Bahcall~N~A, Ostriker~J~P, Perlmutter~S and Steinhardt~P~J 1999
  {\it Science} {\bf 284} 1481
  [arXiv:astro-ph/9906463]
  
\bibitem{Sakharov}
Sakharov~A~D 1967
  {\it Pisma Zh.\ Eksp.\ Teor.\ Fiz.} {\bf 5} 32

\bibitem{EHNOS}
Ellis~J~R, Hagelin~J~S, Nanopoulos~D~V, Olive~K~A and Srednicki~M 1984
  {\it Nucl.\ Phys.}  B {\bf 238} 453
  
\bibitem{axion}
Peccei~R~D and Quinn~H~R 1977
  {\it Phys.\ Rev.\ Lett.}  {\bf 38} 1440;
  Weinberg~S 1978
  {\it Phys.\ Rev.\ Lett.}  {\bf 40} 223;
Wilczek~F 1978
  {\it Phys.\ Rev.\ Lett.} {\bf 40} 279
  
\bibitem{Evans}
Evans~L, talk at this conference: see also {\tt http://lhc.web.cern.ch/lhc/}

\bibitem{Higgs}
Higgs~P~W 1964
  {\it Phys.\ Lett.}  {\bf 12} 132;
  Englert~F and Brout~R 1964
  {\it Phys.\ Rev.\ Lett.}  {\bf 13} 321

\bibitem{Higgsless}
Ozcan~V~E
  arXiv:0710.2786 [hep-ex]
  
\bibitem{Tevatron}
Heinemann~B, talk at this meeting;
Duperrin~A for the CDF and D0 Collaborations
  arXiv:0710.4265 [hep-ex]

\bibitem{POFPA}
Blondel~A, Camilleri~L, Ceccucci~A, Ellis~J~R, Lindroos~M, Mangano~M and Rolandi~G
{\it Physics opportunities with future proton accelerators at CERN}
  arXiv:hep-ph/0609102

\bibitem{Cheng}
Cheng~H~C
  arXiv:0710.3407 [hep-ph]
  
\bibitem{Chanowitz}
Chanowitz~M~S 2002
  {\it Phys.\ Rev.}  D {\bf 66} 073002
  [arXiv:hep-ph/0207123]

\bibitem{Barbieri}
Barbieri~R and Strumia~A
  arXiv:hep-ph/0007265

\bibitem{LHiggs}
Arkani-Hamed~N, Cohen~A~G and Georgi~H 2001
  {\it Phys.\ Lett.}  B {\bf 513} 232
  [arXiv:hep-ph/0105239]
  
\bibitem{boring}
Espinosa~J~R, Giudice~G~F and Riotto~A
  arXiv:0710.2484 [hep-ph];
  and references therein
  
\bibitem{Planck}
Planck Collaboration
  arXiv:astro-ph/0604069
  
\bibitem{KKMR}
Kinney~W~H, Kolb~E~W, Melchiorri~A and Riotto~A 2006
  {\it Phys.\ Rev.}  D {\bf 74} 023502
  [arXiv:astro-ph/0605338]

\bibitem{snuinf}
Murayama~M, Suzuki~H, Yanagida~T and Yokoyama~J 1993
  Phys.\ Rev.\ Lett.\  {\bf 70} 1912
  
\bibitem{ERY}
Ellis~J~R, Raidal~M and Yanagida~T 2004
  Phys.\ Lett.\  B {\bf 581}  9
  [arXiv:hep-ph/0303242]
  
\bibitem{hierarchy}
Maiani~L 1979 {\it All You Need To Know About The Higgs Boson} in Proceedings
of the Gif-sur-Yvette Summer School On Particle Physics pp.1-52; 't~Hooft~G 1979 in {\it
Recent Developments in Gauge Theories} Proceedings of the NATO Advanced Study
Institute, Carg{\`e}se eds. 't~Hooft~G {\it et al} (Plenum Press, NY);
Witten~E 1981
  {\it Phys.\ Lett.} B {\bf 105} 267

\bibitem{GUTs}
Ellis~J~R, Kelley~S and Nanopoulos~D~V 1990
  {\it Phys.\ Lett.} B {\bf 249} 441 and 1991
  {\it Phys.\ Lett.} B {\bf 260} 131;
Amaldi~U, de Boer~W and Furstenau~H 1991
  {\it Phys.\ Lett.} B {\bf 260} 447;
  Giunti~C, Kim~C~W and Lee~U~W 1991
  {\it Mod.\ Phys.\ Lett.} A {\bf 6} 1745

\bibitem{ENOS}
See, for example:
Ellis~J~R, Nanopoulos~D~V, Olive~K~A and Santoso~Y 2006
  {\it Phys.\ Lett.}  B {\bf 633} 583
  [arXiv:hep-ph/0509331];
  Ref.~ \cite{EHOWW};
  Buchmueller~O {\it et al}
  arXiv:0707.3447 [hep-ph]
  
\bibitem{Clowe}
Bradac~M {\it et al} 2006
  {\it Astrophys.\ J.}  {\bf 652} 937
  [arXiv:astro-ph/0608408]
  
\bibitem{Jee}
Jee~M~J {\it et al}
  arXiv:0705.2171 [astro-ph]
  
\bibitem{scaffolding}
Massey~R {\it et al} 2007
  {\it Nature} {\bf 445} 286
  [arXiv:astro-ph/0701594]
  
\bibitem{BNL}
Brown~H~N {\it et al}  [Muon g-2 Collaboration] 2001
  {\it Phys.\ Rev.\ Lett.}  {\bf 86} 2227
  [arXiv:hep-ex/0102017]
  
\bibitem{EOSS}
Ellis~J~R, Olive~K~A, Santoso~Y and Spanos~V~C 2003
  {\it Phys.\ Lett.}  B {\bf 565} 176
  [arXiv:hep-ph/0303043];
  see also:
  Olive~K~A
  arXiv:0709.3303 [hep-ph]

\bibitem{NLSP}
Ellis~J~R, Olive~K~A, Santoso~Y and Spanos~V~C 2004
  Phys.\ Lett.\  B {\bf 603}  51
  [arXiv:hep-ph/0408118]
 
\bibitem{EHOWW}
Ellis~J~R, Heinemeyer~S, Olive~K~A, Weber~A~M and G.~Weiglein 2007
  {\it JHEP} {\bf 0708} 083
  [arXiv:0706.0652 [hep-ph]
  
\bibitem{pQCD}
See, for example:
Cachazo~F, Svrcek~P and Witten~E 2004
  {\it JHEP} {\bf 0409} 006
  [arXiv:hep-th/0403047]

\bibitem{QGP}
See, for example:
Kovtun~P, Son~D~T and Starinets~A~O 2005
  {\it Phys.\ Rev.\ Lett.}  {\bf 94} 111601
  [arXiv:hep-th/0405231]

\bibitem{landscape}
Susskind~L
  arXiv:hep-th/0302219

\bibitem{nosusy}
Arkani-Hamed~N and Dimopoulos~S 2005
  {\it JHEP} {\bf 0506} 073
  [arXiv:hep-th/0405159];
  Giudice~G~F and Romanino~A 2004
  {\it Nucl.\ Phys.}  B {\bf 699} 65
  [Erratum-ibid.\  2005 B {\bf 706} 65]
  [arXiv:hep-ph/0406088]
  
\bibitem{Witten}
Witten~E
  arXiv:hep-th/0106109
  
\bibitem{EMN}
Ellis~J~R, Mavromatos~N~E and Nanopoulos~D~V 1992
  {\it Phys.\ Lett.}  B {\bf 293}  37
  [arXiv:hep-th/9207103]
  
\bibitem{GUTsize}
Witten~E 1996
  {\it Nucl.\ Phys.}  B {\bf 471} 135
  [arXiv:hep-th/9602070]
  
\bibitem{Ignatios}
Antoniadis~A 1990
  {\it Phys.\ Lett.}  B {\bf 246} 377
  
\bibitem{ADD}
Antoniadis~A, Arkani-Hamed~N, Dimopoulos~S and Dvali~G~R 1998
  Phys.\ Lett.\  B {\bf 436} 257
  [arXiv:hep-ph/9804398]
  
\bibitem{RS}
Randall~L and Sundrum~R 1999
  {\it Phys.\ Rev.\ Lett.}  {\bf 83} 3370
  [arXiv:hep-ph/9905221]
  
\end{thebibliography}
\end{document}